\documentclass[12pt]{iopart}

\usepackage{epsfig}
\usepackage{iopams}

\usepackage{graphicx}
\usepackage{color}
\usepackage{amsthm}
\usepackage{bm}
\usepackage{layout}
\usepackage{float}

\usepackage{amsfonts}
\usepackage{amssymb}%
\begin{document}

\newcommand{\ket}[1]{\left | #1 \right\rangle}
\newcommand{\bra}[1]{\left \langle #1 \right |}
\newcommand{\half}{\mbox{$\textstyle \frac{1}{2}$}}
\newcommand{\smallfrac}[2][1]{\mbox{$\textstyle \frac{#1}{#2}$}}
\newcommand{\braket}[2]{\left\langle #1|#2\right\rangle}
\newcommand{\proj}[1]{\ket{#1}\bra{#1}}
\renewcommand{\epsilon}{\varepsilon}

\title[Density cubes and higher-order interference theories]{Density cubes and higher-order interference theories}

\author{B Daki\'{c}$^{1,2}$,
T Paterek$^{2,3}$
and {\v C} Brukner$^{1,4}$}

\address{
$^1$ Vienna Center for Quantum Science and Technology (VCQ), Faculty of Physics, University of Vienna,
Boltzmanngasse 5, A-1090 Vienna, Austria \\
$^2$ Centre for Quantum Technologies, National University of Singapore, 3 Science Drive 2, Singapore 117543\\
$^3$ School of Physical and Mathematical Sciences, Nanyang Technological University, , 21 Nanyang Link, Singapore 637371 \\
$^4$ Institute of Quantum Optics and Quantum Information, Austrian Academy of Sciences, Boltzmanngasse 3, A-1090 Vienna, Austria}

\date{\today}

\begin{abstract}
Can quantum theory be seen as a special case of a more general probabilistic theory, similarly as classical theory is a special case of the quantum one?
We study here the class of generalized probabilistic theories defined by the order of interference they exhibit as proposed by Sorkin.
A simple operational argument shows that the theories require higher-order tensors as a representation of physical states.
For the third-order interference we derive an explicit theory of ``density cubes'' and show that quantum theory, i.e. theory of density matrices, is naturally embedded in it.
We derive the genuine non-quantum class of states and non-trivial dynamics for the case of three-level system and show how one can construct the states of higher dimensions.
Additionally to genuine third-order interference, the density cubes are shown to violate the Leggett-Garg inequality beyond the quantum Tsirelson bound for temporal correlations.
\end{abstract}

\maketitle

\section{Introduction}

In the past researchers were often deeply convinced of the absolute validity of the ruling set of theories, and yet the set was later inevitably replaced by a more fundamental one of which the old one has remained a special case.
In this respect it seems of supreme importance to keep performing dedicated tests of foundations of quantum physics with the goal of possibly finding a cue for deviations from what we presently expect.
A vast majority of the tests performed to date contrast quantum mechanical predictions with the predictions of those theories that preserve one or other notion of classical physics intact.
Examples are hidden-variable theories~\cite{KS,Bell,Leggett}, non-linear modifications of the Schr\"{o}dinger equation~\cite{Birula,Shimony,Shull,Gaehler} or the collapse models~\cite{GWR,Karoli1,Diosi,Penrose,Pearle}.
Hidden variables pre-assign definite values to outcomes of unperformed measurements, non-linear Schr\"{o}dinger equations allow solutions with localized wave-packets to resemble classical trajectories and collapse models restore macrorealism by suppressing superpositions between macroscopically distinct states.
Judging from historical experience, however, it seems very unlikely that a post-quantum theory will be based on pre-quantum concepts.
In contrast, one might expect that it will break not only postulates of classical but also quantum theory~\cite{PhysToday}.

Quantum mechanics can be seen as particular theory of the class of generalized probabilistic theories~\cite{PRboxes,Karol,Barrett,Hardy,Steeg,PDB,Sorkin}. These theories share with quantum mechanics its non-classical features such as randomness of individual results, the impossibility of copying unknown states~\cite{NO_CLONING,Barnum07}, violation of Bell's inequalities~\cite{Bell}, uncertainty relations~\cite{Steeg,PDB} or interference~\cite{Sorkin}.
Recent progress in reconstructions of quantum  formalism give a variety of choices for postulates on which quantum theory can be singled out from the class of generalized probabilistic theories~\cite{Fivel, Hardy, DakicBrukner, MassanesMueler, Rau, Dariano1}.

Quantum interference is standardly explained through the double-slit experiment, where one combines superposition principle and Born rule to derive the ``interference term'' -- a quantity that vanishes in all classical experiments.
If one considers multi-slit interference experiments there is a very natural hierarchy of probabilistic theories due to Sorkin~\cite{Sorkin}.
The hierarchy is described by the \emph{order of interference} $I_k$ ($k = 1, 2, 3,\dots$) defined by the outcome probabilities in $k$-slit experiment.
One may consider $I_k$ as a measure of genuine coherence between $k$ slits.
Sorkin showed that quantum mechanics exhibits only two-slit interference, but no genuine three-slit or higher-order interferences.
This demonstrates that a theory that exhibits, for example, genuine three-slit interference $I_3$ is essentially a non-quantum theory though no explicit such theory is known.
Recently, an experiment has been performed that puts a bound to third-order interference term to less than $10^{-2}$ of the regular second order interference~\cite{Sinha}
and further experiments are planned to improve the bound~\cite{Immo}.

In this work we give operational arguments why the theory that exhibits the $k$th-order interference describes physical states by tensors with $k$ indices.
For example, classical probability theory as a first-order theory represents states by a (probability) vector (one-index tensor) and quantum theory by a density matrix.
Similarly, for a third-order theory one needs an object with three indices and we call it a \emph{density cube}.
We develop a theory of density cubes and show that it contains quantum theory as a subset, the same way quantum theory contains classical theory as a subset.
We derive a class of non-quantum states and show that there exists non-trivial dynamics that maps between quantum and non-quantum states.
All this allows creation, manipulation and tomography of density cubes as well as violation of the Leggett-Garg inequality~\cite{LG} beyond the quantum Tsirelson bound for temporal correlations~\cite{Tsirelson, Fritz}.
It was recently shown that the absence of third-order interference implies the validity of Tsirelson's bound for spatial correlations for a broad class of probabilistic theories~\cite{Niestegge}.

\section{Higher-order interference theories}\label{HO theories}

\begin{figure}\centering
\includegraphics[width=10cm]{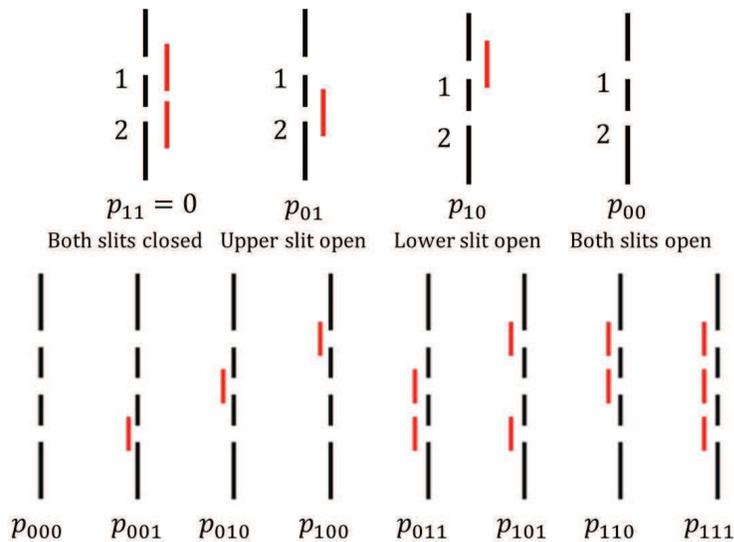}
\caption{(Up) The set of four experimental set-ups with two, one and no slits opened for probing if the theory has a nonvanishing second-order interference
(such as complex, quaternion or octonion quantum mechanics do). (Down) The complete series of ``single-slit'' and ``double-slit'' experimental set-ups
to probe if the theory in test shows third-order interference.}
\label{Figure1}
\end{figure}

In Ref.~\cite{Sorkin} Sorkin suggested a classification of theories according to the order of interference the theory exhibits.
Roughly speaking, the order indicates how much the calculus of the predicted probabilities in the theory deviates from the one in classical physics.
To demonstrate the interference phenomenon we first consider the double-slit experiment, as shown in Fig.~\ref{Figure1}, and a series of set-ups in which each slit is either open or closed.
We distinguish four situations: both slits are open, either one of them is closed and both are closed.
The four physical situations are labeled as $00,01,10,11$, where, e.g. $01$ denotes the scenario with the lower slit blocked.
The ``non-classicality'' is measured via the interference term:
\begin{eqnarray}
I_{12} & = & p_{00}-p_{01}-p_{10}+p_{11},
\label{I2}
\end{eqnarray}
where $p_{ij}$ is the conditional probability to find the particle at a certain point on the observation screen given that situation $ij$ is realized.
Of course, $p_{11}$ always vanishes bringing no contribution to the value of $I_{12}$ and we introduce it only for symmetry reasons and future generalization.

Classical theory of bullets belongs to the lowest class in this hierarchy because $I_{12}=0$.
Quantum theory is an example of a theory for which $I_{12}$ does not vanish, and we call this feature the second-order interference.

Consider now a triple-slit experiment.
One could propose to measure the third-order interference by the quantity $I_{123}=p_{000}-p_{011}-p_{101}-p_{110}$, where $p_{011}$ is the
conditional probability to detect the particle at a certain point on the observation screen, if the upper slit is open and the middle and the lower
slits are blocked (see lower panel of Fig.~\ref{Figure1}).
Note, however, that $I_{123}\neq 0$ can solely be due to the non-vanishing second-order interference terms. In order to quantify genuine third-order interference one thus
needs to subtract all the two-order interference terms
\begin{eqnarray}
I_{12} & = & p_{001}-p_{011}-p_{101},\nonumber \\
I_{13} & = & p_{010}-p_{011}-p_{110},\\
I_{23} & = & p_{100}-p_{101}-p_{110}, \nonumber
\end{eqnarray}
from $I_{123}$. Therefore, as a measure of genuine third-order interference one introduces
\begin{eqnarray}
I_{123} & = & \sum_{i, j, k=0}^{1}(-1)^{i+j+k}p_{ijk},
\end{eqnarray}
where again the last term $p_{111}$ always vanishes.
Straightforward calculation shows that
$I_{123}=0$ in quantum theory, regardless of the Hilbert space dimension of the system and the type of measurement.
In fact all theories that are represented by Jordan algebras~\cite{JNW} have vanishing the third-order interference as well.
Examples contain quantum mechanics based on complex, quaternion or octonion probability amplitudes.
The theories for which the third-order interference term is zero have been characterized as those in which it is possible to fully determine the state, i.e. to perform state tomography, via a complete set of single-slit and double-slit experiments~\cite{Ududec}.

Generally speaking, the Sorkin's quantity $I_{12...k}$ measuring the genuine $k$th-order interference is a sum of interference terms for all combinations of open slits where the terms involving $k-j$ slits, for odd $j$, enter with a minus sign and for even $j$, with a plus sign. If $I_{12...k} = 0$, then all $I_{12...l}$, with $l>k$, also vanish~\cite{Sorkin}. The level of a theory is the highest $k$ for which the Sorkin's quantity $I_{12...k}$ does not vanish.

\section{Theory of density cubes}

We will now develop an explicit theory that belongs to the Sorkin's class of theories with non-vanishing third-order interference. The theory contains quantum theory as a subset.

The fact that $I_{123}=0$ in quantum mechanics can be traced back to the description of state in terms of density matrix elements $\rho_{ij}$ that link coherently at most \textit{two} different states $i$ and $j$.
It is therefore natural to assume that if a theory should allow for $I_{123} \neq 0$, then the description of the state in the theory should involve elements that link \textit{three} different states $i$, $j$, $k$, i.e. it should have elements of the form $\rho_{ijk}$. We consider a framework in which the description of the state is given by the tensor with elements $\rho_{ijk}$, and we call it a density cube.

We follow the analogy to the quantum case in order to define the basic ingredients of the theory.
To every measurement outcome we associate a density cube with, in general, complex entries $\rho_{ijk}$.
The element $\rho_{iii}$ is chosen to be real and gives the probability for the outcome $i$, therefore $\sum_i\rho_{iii}=1$ and $\rho_{iii}\geq0$. The Born rule in quantum mechanics reads $p=\Tr\rho^{\dagger}\sigma=\rho^{\ast}_{ij}\sigma_{ij}$, where we adopt Einstein's summation rule. In a similar manner we define
\begin{eqnarray}
p & = & (\rho,\sigma)=\rho^{\ast}_{ijk}\sigma_{ijk}.
\label{BR}
\end{eqnarray}
Here $p$ denotes the probability, whereas $\rho$ and $\sigma$, are the state and element of a projective measurement, respectively. Note, that we follow the analogy to quantum theory where there exists a compete identification between states and measurements. Therefore, $\rho$ and $\sigma$ belong essentially to the same set. To ensure that $p$ is a real number, we put the constraint $\rho^{\ast}_{ijk}\sigma_{ijk}=\sigma^{\ast}_{ijk}\rho_{ijk}$.
In the quantum case $p\in\mathbb{R}$ is provided by the fact that $\rho_{ij}$ is a Hermitian matrix, hence $\rho_{ij}^{\ast}=\rho_{ji}$. Similarly, we
expect $\rho_{ijk}^{\ast}=\rho_{\pi(ijk)}$, where $\pi(ijk)$ is some permutation of indices $ijk$.
The condition $(\rho_{ijk}^{\ast})^{\ast}=\rho_{ijk}$ implies that $\pi$ is the index transposition.
Accordingly, we call the cubes Hermitian if exchanging two indices gives a complex conjugated element.
As in the case of Hermitian matrices, Hermitian cubes form a real vector space
with the inner product given by~(\ref{BR}). Indeed, $(\rho,\sigma)^{\ast}=\sigma^{\ast}_{ijk}\rho_{ijk}=\sigma_{jik}\rho_{ijk}=\rho_{jik}\sigma_{ijk}=\rho^{\ast}_{ijk}\sigma_{ijk}=(\rho,\sigma)$.
For $\rho_{ijk}$ being a pure state we expect $(\rho,\rho)=1$.

\subsection{Two-level system}
For a system with two distinguishable outcomes, the hermiticity constraint together with the normalization condition reads:
\begin{eqnarray}
\rho_{112}^{\ast}&=&\rho_{112}=\rho_{121}=\rho_{211}\in\mathbb{R},\nonumber \\
\rho_{122}^{\ast}&=&\rho_{122}=\rho_{212}=\rho_{221}\in\mathbb{R},\nonumber \\
\rho_{111}&+&\rho_{222}=1,\nonumber\\
\rho_{iii}&\geq&0,~i=1,2.
\end{eqnarray}
There are three independent real parameters here, e.g. $\rho_{111}$, $\rho_{112}$, and $\rho_{122}$, and we can write a density cube as a list of two matrices
\begin{eqnarray}
\rho & = & \{\left(
         \begin{array}{cc}
           \rho_{111} & \rho_{112} \\
           \rho_{112} & \rho_{122} \\
         \end{array}
       \right),\left(
                 \begin{array}{cc}
                   \rho_{112} & \rho_{122} \\
                   \rho_{122} & 1-\rho_{111} \\
                 \end{array}
               \right)
\},
\end{eqnarray}
where the $i$th list element is the matrix $\rho_{ijk}$. We set $\rho_{112}=\frac{x_1}{\sqrt{6}}$, $\rho_{122}=\frac{x_2}{\sqrt{6}}$, and $\rho_{111}=(1+x_3)/2$. In this parametrization the normalization condition for pure states $(\rho,\rho)=1$ is equivalent to $x_1^2+x_2^2+x_3^2=1$, therefore the set of pure states is isomorphic to the Bloch sphere. We can define the four Pauli cubes
\begin{eqnarray}
\sigma_0&=&\{\left(
         \begin{array}{cc}
           1 & 0 \\
           0 & 0 \\
         \end{array}
       \right),\left(
                 \begin{array}{cc}
                   0 & 0 \\
                   0 & 1 \\
                 \end{array}
               \right)
\},\nonumber \\
\sigma_1&=&\sqrt{\frac{2}{3}}\{\left(
         \begin{array}{cc}
           0 & 1 \\
           1 & 0 \\
         \end{array}
       \right),\left(
                 \begin{array}{cc}
                   1 & 0 \\
                   0 & 0 \\
                 \end{array}
               \right)
\},\nonumber\\
\sigma_2&=&\sqrt{\frac{2}{3}}\{\left(
         \begin{array}{cc}
           0 & 0 \\
           0 & 1 \\
         \end{array}
       \right),\left(
                 \begin{array}{cc}
                   0 & 1 \\
                   1 & 0 \\
                 \end{array}
               \right)
\}, \nonumber \\
\sigma_3&=&\{\left(
         \begin{array}{cc}
           1 & 0 \\
           0 & 0 \\
         \end{array}
       \right),\left(
                 \begin{array}{cc}
                   0 & 0 \\
                   0 & -1 \\
                 \end{array}
               \right)
\}.
\end{eqnarray}
They span the set of Hermitian cubes and we can write $\rho=(\sigma_0+\vec{x}\cdot\vec{\sigma})/2$, where $\vec{\sigma}\equiv (\sigma_1,\sigma_2,\sigma_3)$. Hence, the set of density cubes for a two-level system is equivalent to the set of states of a qubit.
This is intuitively expected as the departure from quantum theory should rather be seen if at least three states are allowed due to genuine third-order interference.

\subsection{Three-level system}

For the case of three-level system the hermiticity condition together with normalization reads:
\begin{eqnarray}
\rho_{112}&=&\rho_{112}^{\ast}=\rho_{121}=\rho_{211},\nonumber\\
\rho_{122}&=&\rho_{122}^{\ast}=\rho_{212}=\rho_{221},\nonumber\\
\rho_{113}&=&\rho_{113}^{\ast}=\rho_{131}=\rho_{311},\nonumber\\
\rho_{133}&=&\rho_{133}^{\ast}=\rho_{313}=\rho_{331},\nonumber\\
\rho_{223}&=&\rho_{223}^{\ast}=\rho_{232}=\rho_{322},\nonumber\\
\rho_{233}&=&\rho_{233}^{\ast}=\rho_{323}=\rho_{332},\nonumber\\
\rho_{123}&=&\rho_{312}=\rho_{231}=\rho_{213}^{\ast}=\rho_{321}^{\ast}=\rho_{132}^{\ast}\nonumber\\
\rho_{111}&+&\rho_{222}+\rho_{333}=1,\nonumber\\
\rho_{iii}&\geq&0,~i=1,2,3.
\end{eqnarray}
Hence we have in total ten real parameters: $\rho_{iii},\rho_{iij}\in\mathbb{R}$ and $z=\rho_{ijk}\in\mathbb{C}$ with all three different indices, which is two real parameters more (one complex parameter)
than what is required to describe a general state of a quantum mechanical three-level system, a qutrit. The parameter $z$ brings the crucial difference between the density matrix and
the density cube.
If $z=0$ the set of cube states is equivalent to the qutrit state space.
Indeed, we can map any density matrix $\rho_{ij}$ to the density cube $\rho_{ijk}$ with $z=0$ in the following way
\begin{eqnarray}
\rho_{iii} & = & \rho_{ii}, \quad \rho_{iij} = \sqrt{\frac{2}{3}}\mathrm{Re}\rho_{ij}, \quad \rho_{ijj} = \sqrt{\frac{2}{3}}\mathrm{Im}\rho_{ij}\mbox{ } \mbox{  for } i<j ,\label{QtoC}
\end{eqnarray}
and vice versa. This mapping preserves the inner product, thus we have a complete (physical) identification between the two sets.

The analysis becomes more intriguing when we consider some genuine non-quantum $\rho_{ijk}$ with $z\neq0$. Positivity condition for two density cubes
$\rho$ and $\sigma$ has to be fulfilled, hence $(\rho,\sigma) \geq 0$. The complete characterization of the set of density cubes remains an open problem,
and here we present one explicit class of states that is in agreement with the positivity condition and extends standard quantum theory.
Consider the pure states
\begin{eqnarray}\label{example}
\rho_1&\!=\!&\{\left(
          \begin{array}{ccc}
            0 & 0 & 0 \\
            0 & 0 & \frac{1}{2\sqrt3} \\
            0 & \frac{1}{2\sqrt3} & 0 \\
          \end{array}
        \right),\left(
          \begin{array}{ccc}
            0 & 0 & \frac{1}{2\sqrt3} \\
            0 & \frac{1}{2} & 0 \\
            \frac{1}{2\sqrt3} & 0 & 0 \\
          \end{array}
        \right),\left(
          \begin{array}{ccc}
            0 & \frac{1}{2\sqrt3} & 0 \\
            \frac{1}{2\sqrt3} & 0 & 0 \\
            0 & 0 & \frac{1}{2} \\
          \end{array}
        \right)
\},\nonumber\\
\rho_2&\!=\!&\{\left(
          \begin{array}{ccc}
            \frac{1}{2} & 0 & 0 \\
            0 & 0 & \frac{\omega}{2\sqrt3} \\
            0 & \frac{\omega^{\ast}}{2\sqrt3} & 0 \\
          \end{array}
        \right),\left(
          \begin{array}{ccc}
            0 & 0 & \frac{\omega^{\ast}}{2\sqrt3} \\
            0 & 0 & 0 \\
            \frac{\omega}{2\sqrt3} & 0 & 0 \\
          \end{array}
        \right),\left(
          \begin{array}{ccc}
            0 & \frac{\omega}{2\sqrt3} & 0 \\
            \frac{\omega^{\ast}}{2\sqrt3} & 0 & 0 \\
            0 & 0 & \frac{1}{2} \\
          \end{array}
        \right)
\},\\
\rho_3&\!=&\!\{\left(
          \begin{array}{ccc}
            \frac{1}{2} & 0 & 0 \\
            0 & 0 & \frac{\omega^{\ast}}{2\sqrt3} \\
            0 & \frac{\omega}{2\sqrt3} & 0 \\
          \end{array}
        \right),\left(
          \begin{array}{ccc}
            0 & 0 & \frac{\omega}{2\sqrt3} \\
            0 & \frac{1}{2} & 0 \\
            \frac{\omega^{\ast}}{2\sqrt3} & 0 & 0 \\
          \end{array}
        \right),\left(
          \begin{array}{ccc}
            0 & \frac{\omega^{\ast}}{2\sqrt3} & 0 \\
            \frac{\omega}{2\sqrt3} & 0 & 0 \\
            0 & 0 & 0 \\
          \end{array}
        \right)
\}, \nonumber
\end{eqnarray}
where $\omega=e^{i2\pi/3}$.
They form a ``basis set'' $(\rho_i,\rho_j) = \delta_{ij}$. When we refer to the ``basis set'' of density cubes, we mean the set of normalized, mutually orthogonal density cubes. Operationally, it means that states $\rho_i$ can be distinguished in a single-shot experiment. Note, that this set is not a complete basis in the vector space of Hermitian cubes.
The quantum parts of these states correspond to density matrices $\frac{1}{2}(\hat 1-\ket{e_i}\bra{e_i})$, where $\ket{e_i}$ are the states of the standard (computational) basis, e.g. $\ket{e_1}=(1,0,0)^{\mathrm{T}}$.
This way, to each pure quantum state $\ket{\psi}=(c_1,c_2,c_3)^{\mathrm{T}}$, we can associate one of three density cubes $\rho^{(n)}(\psi)$ with the following elements
\begin{eqnarray}
\rho^{(n)}_{iij}&=&-\frac{1}{\sqrt6}\mathrm{Re}(c_i^{\ast}c_j) \nonumber \\
\rho^{(n)}_{ijj}&=&-\frac{1}{\sqrt6}\mathrm{Im}(c_i^{\ast}c_j),~~i<j\\
\rho^{(n)}_{iii}&=&\frac{1}{2}(1-|c_i|^2),~~i=1,2,3 \nonumber\\
\rho^{(n)}_{123}&=&\frac{\omega^n}{2\sqrt{3}}, \nonumber
\end{eqnarray}
where $n=1,2,3$. Straightforward verification shows $(\rho^{(n)}(\psi),\rho^{(m)}(\phi))=\frac{1}{4}(1+|\braket{\psi}{\phi}|^2)+\frac{1}{2}\cos\frac{2\pi(n-m)}{3}\geq0$, hence the positivity is preserved. Furthermore, one can verify that there are at most three mutually orthogonal density cubes, within the class of states $\rho^{(n)}(\psi)$.

\subsection{Non-trivial dynamics}\label{Non-trivial dynamics}

We derived the class of ``non-trivial'' density cubes, and the next step is to give an example of genuine ``non-quantum'' transformation (evolution).
We define the unitary (norm preserving) transformation $T: \beta_0 \to \beta$, where $\beta_0=\{e_1,e_2,e_3\}$ is the standard basis set, i.e. $[e_n]_{ijk}=\delta_{in}\delta_{jn}\delta_{kn}$ and $\beta=\{\rho_1,\rho_2,\rho_3\}$ is the basis set defined in Eq.~(\ref{example}). Rather than working directly with the whole set of Hermitian cubes we define the following subspace spanned by the (sub)basis
\begin{eqnarray}
E^{(n)}_{ijk}&=&\delta_{in}\delta_{jn}\delta_{kn},~~n=1,2,3, \nonumber \\
E^{(4)}&=&\frac{1}{\sqrt3}\{\left(
          \begin{array}{ccc}
            0 & 0 & 0 \\
            0 & 0 & 1 \\
            0 & 0 & 0 \\
          \end{array}
        \right),\left(
          \begin{array}{ccc}
            0 & 0 & 0 \\
            0 & 0 & 0 \\
            1 & 0 & 0 \\
          \end{array}
        \right),\left(
          \begin{array}{ccc}
            0 & 1 & 0 \\
            0 & 0 & 0 \\
            0 & 0 & 0 \\
          \end{array}
        \right)
\},\\
E^{(5)}&=&\frac{1}{\sqrt3}\{\left(
          \begin{array}{ccc}
            0 & 0 & 0 \\
            0 & 0 & 0 \\
            0 & 1 & 0 \\
          \end{array}
        \right),\left(
          \begin{array}{ccc}
            0 & 0 & 1 \\
            0 & 0 & 0 \\
            0 & 0 & 0 \\
          \end{array}
        \right),\left(
          \begin{array}{ccc}
            0 & 0 & 0 \\
            1 & 0 & 0 \\
            0 & 0 & 0 \\
          \end{array}
        \right)
\}. \nonumber
\end{eqnarray}
Note that this is not a complete basis within the vector space of Hermitian cubes. However, we assume that $T$ keeps the subspace spanned by the subbasis invariant, that is if
$\rho\in\mathrm{span}\{E^{(n)}\}$ ($n=1,...,5$), then necessarily $T\rho\in\mathrm{span}\{E^{(n)}\}$. Recall that $T$ has been defined as a unitary transformation. Therefore, within this invariant subspace, $T$ is represented by a $5\times5$ unitary matrix. The basis sets $\beta_0$ and $\beta$ are represented by the vectors
%\begin{eqnarray}
%&&e_1=(1,0,0,0,0)^{\mathrm{T}},~~e_2=(0,1,0,0,0)^{\mathrm{T}},~~e_3=(0,0,1,0,0)^{\mathrm{T}},\\
%&&\rho_1=\frac{1}{2}(0,1,1,1,1)^{\mathrm{T}},~~\rho_2=\frac{1}{2}(1,0,1,\omega,\omega^{\ast})^{\mathrm{T}},~~e_3=(1,1,0,\omega^{\ast},\omega)^{\mathrm{T}}.
%\end{eqnarray}
\begin{eqnarray}
&&e_1=(1,0,0,0,0)^{\mathrm{T}}, \hspace{0.4cm} \rho_1=\frac{1}{2}(0,1,1,1,1)^{\mathrm{T}},\nonumber\\
&&e_2=(0,1,0,0,0)^{\mathrm{T}}, \hspace{0.4cm} \rho_2=\frac{1}{2}(1,0,1,\omega,\omega^{\ast})^{\mathrm{T}}, \\
&&e_3=(0,0,1,0,0)^{\mathrm{T}}, \hspace{0.4cm} \rho_3=\frac{1}{2}(1,1,0,\omega^{\ast},\omega)^{\mathrm{T}},\nonumber
\end{eqnarray}
respectively. The condition $Te_i=\rho_i$ leads to the following matrix
\begin{eqnarray}
T & = & \frac{1}{2}\left(
    \begin{array}{ccccc}
      0 & 1 & 1 & a_1 & b_1 \\
      1 & 0 & 1 & a_2 & b_2 \\
      1 & 1 & 0 & a_3 & b_3 \\
      1 & \omega & \omega^{\ast} & a_4 & b_4 \\
      1 & \omega^{\ast} & \omega & a_5 & b_5 \\
    \end{array}
  \right),
\end{eqnarray}
where $a_i$ and $b_i$ are unknown coefficients. All the columns of the matrix $T$ have to be orthogonal vectors, hence
\begin{eqnarray}\label{cond1}
a_2+a_3+a_4+a_5 & = &0,\nonumber\\
a_1+a_3+\omega^{\ast}a_4+\omega a_5 & = &0,\nonumber\\
a_1+a_2+\omega a_4+\omega^{\ast} a_5 & = & 0,\\
b_2+b_3+b_4+b_5 & = & 0,\nonumber\\
b_1+b_3+\omega^{\ast}b_4+\omega b_5 & = & 0,\nonumber\\\label{cond-1}
b_1+b_2+\omega b_4+\omega^{\ast} b_5 & = & 0, \nonumber
\end{eqnarray}
and $\sum_{i=1}^5a_i^{\ast}b_i=0$.
The set of solutions of the  equations above is manifold. Here we present one particular solution 

%If we apply $T$ to the state $\rho_1$ we obtain
%\begin{eqnarray}
%T\rho_1 & = & \frac{1}{4}\left(
%                     \begin{array}{c}
%                       2+a_1+b_1 \\
%                       1+a_2+b_2 \\
%                       1+a_3+b_3 \\
%                       -1+a_4+b_4 \\
%                       -1+a_5+b_5 \\
%                     \end{array}
%                   \right),
%\end{eqnarray}
%where we used $\omega+\omega^{\ast}=-1$. Keeping in mind that Eqs.~(\ref{cond1})-(\ref{cond-1}) have to hold, straightforward calculation gives
%\begin{eqnarray}
%(\rho_1,T\rho_1) & = & 0, ~~ (\rho_2,T\rho_1)=\frac{1}{2}, ~~ (\rho_3,T\rho_1)=\frac{1}{2}.
%\end{eqnarray}
%The only pure state that is orthogonal to $\rho_1$ and  has the non-zero overlap with $\rho_2$ and $\rho_3$ is the state $e_1$, therefore $T\rho_1=e_1$. Similarly we can derive $T\rho_i=e_i$, which implies $T$ being an involution $T^2=\hat 1$. One thus has
\begin{eqnarray}
T & = & \frac{1}{2}\left(
    \begin{array}{ccccc}
      0 & 1 & 1 & 1 & 1 \\
      1 & 0 & 1 & \omega^{\ast} & \omega \\
      1 & 1 & 0 & \omega & \omega^{\ast} \\
      1 & \omega & \omega^{\ast} & 1 & 0 \\
      1 & \omega^{\ast} & \omega & 0 & 1 \\
    \end{array}
  \right). \label{T}
\end{eqnarray}
It is easy to verify $T$ being an involution $T^2=\hat 1$. Transformation $T$ is distinct from any unitary transformations for a qutrit. To see this, suppose that some unitary $U$ maps the vector of the
standard basis $\ket{e_i}$ such that $|\langle e_j | U | e_i \rangle |^2=(1-\delta_{ij})/2$. Note that matrix $U$ has zeros at the main diagonal, whereas all the off-diagonal elements have non-zero value. This implies that its columns cannot form a set of orthogonal vectors, and thus $U$ is not a unitary matrix.

\subsection{Measurement and the state update rule}
When a measurement is performed and an outcome is obtained, our knowledge
about the state of the system changes and its representation must be updated to be in agreement with the new knowledge
acquired in the measurement. In quantum theory, when we refer to a state of a $N$-level system defined by a density matrix $\rho_{ij}$, we implicitly assume that the matrix elements are defined with respect to some basis of orthogonal states $\beta=\{\ket{e_n},n=1\dots N\}$. We call this basis set a measurement or computational basis. If a projective (selective or von Neumann) measurement is performed and outcome $n$ has been observed, the state is updated to a pure state $\rho_{ij}=\delta_{in}\delta_{jn}$, and each outcome occurs with the probability $p_n=\rho_{nn}$. An operational argument for this update rule is that immediate consecutive measurements of the same observable should give the same result. More generally, a non-selective measurement can be performed, i.e. the outcomes are ``merged'' into $d$ non-overlapping slots that correspond to the partition of basis $\beta=\{\beta_1,\beta_2,\dots,\beta_d\}$. The set of outcomes is fully specified by the set of orthogonal projectors $P_s=\sum_{\ket{e_k}\in\beta_s}\ket{e_k}\bra{e_k}$, each of which projects onto subspace spanned by $\beta_s$. In this case, for an outcome $s$ obtained in a measurement, the generalized projection postulate (L\" uders rule~\cite{Luders}) applies
\begin{equation}
\rho\mapsto\frac{1}{p_s}P_s\rho P_s,
\end{equation}
where $p_s=\mathrm{Tr}P_s\rho$. The update rule is such that the elements are either re-normalized $\rho_{ij}\mapsto\frac{1}{p_s}\rho_{ij}$ (for $\ket{e_i},\ket{e_j}\in\beta_s$) or ``destroyed'' $\rho_{ij}\mapsto0$ (for $\ket{e_i}\not\in\beta_s$ or $\ket{e_j}\not\in\beta_s$).
In general case, the measurement can be performed in a non-computational basis $\beta'=U\beta$, where $U$ is a unitary transformation. In such a case, we can simply apply a transformation to the initial state $\rho'=U\rho U^{\dagger}$ and apply the update rule to the matrix $\rho'_{ij}$.

Now we can use analogy to the quantum case to define the L\" uders rule for the theory of density cubes. When we refer to a density cube $\rho_{ijk}$ we implicitly assume that the cube elements are defined with respect to some basis of orthogonal cubes $\beta=\{e_n,n=1\dots N\}$ (for a $N$-level system in general, see Section \ref{HLSys}). If a projective measurement is performed and the outcome $n$ has been observed, the state of system $\rho_{ijk}$ is simply updated to a pure state $e_n$, i.e. $\rho_{ijk}\mapsto\delta_{in}\delta_{jn}\delta_{kn}$, and each outcome occurs with the probability $p_n=\rho_{nnn}$. More generally, a non-selective measurement with set of $d$ outcomes is defined by some partition of computational basis $\beta=\{\beta_1,\beta_2,\dots,\beta_d\}$. The probability to observe the outcome $s$ can by computed using Eq. (\ref{BR}) and is given by $p_s=\sum_{e_k\in\beta_s}(e_k,\rho)$. Similarly to the quantum case, the generalized update rule is such the the cube elements are either re-normalized $\rho_{ijk}\mapsto\frac{1}{p_s}\rho_{ijk}$ (for $e_i,e_j,e_k\in\beta_s$) or ``destroyed'' $\rho_{ijk}\mapsto0$ (for $e_i\not\in\beta_s$ or $e_j\not\in\beta_s$ or $e_k\not\in\beta_s$).

\subsection{Genuine three-slit interference}
\begin{figure}\begin{center}
\includegraphics[width=11cm]{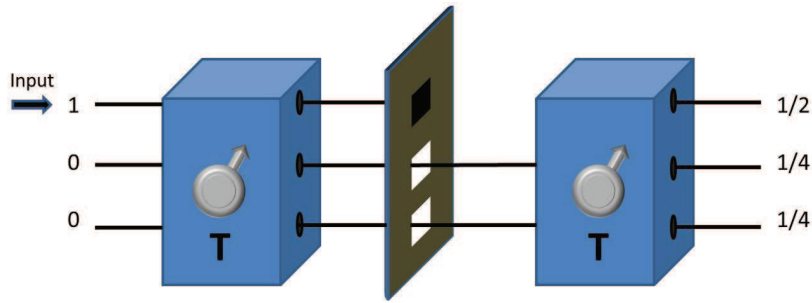}
\caption{Genuine \emph{three-slit} interference.
The interferometer consists of two consecutive transformations $T$ of Eq.~(\ref{T}). The horizontal lines denote three possible paths a system can take.
To compute the third-order interference $I_{123}$, eight experimental runs are distinguished in which different set of slits are open or closed (see Section ~\ref{HO theories}). The setup with the upper slit closed and two lower slits open is shown here. The numbers close to the paths describe the probabilities of finding the system in a particular path at the initial and final stage. The experiment described allows non-trivial third-order interference (see main text).}
\label{FIG123}
\end{center}
\end{figure}
After defining a non-trivial class of states, non-trivial evolution and the measurement update rule, we show how the theory of density cubes allows for a genuine \emph{third-order} interference.

This could be demonstrated either in a triple-slit experiment or its finite-dimensional analogue, a Mach-Zehnder-like interferometer with three possible paths and ``beam-splitters'', as shown in Fig~\ref{FIG123}. In standard quantum mechanics, the role of the first beam-splitter is to prepare a superposition of a system propagating along the three paths, and the second beam-splitter coherently recombines the beams.
%In the analogous triple slit experiment the beam-splitters are replaced with free evolution. At first, the free evolution from a source to the slits essentially prepares a plane wave at the screen with the slits, next the free evolution governs the probability of the system arriving at a given point on the observation screen.
For our purposes, we focus on the Mach-Zehnder three-path interferometer and replace the beam-splitters with transformation $T$.

Consider an interferometer, shown in Fig.~\ref{FIG123}, that consist of two consecutive transformations $T$ of Eq.~(\ref{T}) and a filter in between. We distinguish eight different situations dependent on which set of slits is opened or closed (see Section~\ref{HO theories}). Let us compute in detail the conditional probability $p_{100}$ for situation shown in Fig.~\ref{FIG123} (upper slit closed, two lower slits open). The system is initially prepared in a state $e_1$ (with certainty in the upper path). The probability of finding the system in a state $e_1$ after passing the first transformation device is zero, $[\rho_1]_{111}=0$, therefore, one could expect that the measurement induced by the filter does not affect the state (as it would be the situation in quantum theory). However, the state $\rho_1$ has a non-zero element $[\rho_1]_{123}=\frac{1}{2\sqrt{2}}$, hence the update rule defined in a previous section enforces a ``collapse'' of the state to the equally weighted mixture $\frac{1}{2}e_2+\frac{1}{2}e_3$. Therefore, the conditional probability of finding a system in  a upper path after passing the second transformation device is $p_{100}=\frac{1}{2}$. Similarly one can compute the conditional probabilities $p_{ijk}$ for the remaining of situations. It is easy to check that $I_{123}=\frac{1}{2}$.

We speculate on how a potential experiment could test the genuine triple-slit interference. An experimenter could probe different media that may be implementing a non-quantum transformation $T$. By embedding the three slits in such a ``T-medium'' one arranges a situation where this operation is applied before and after the propagation of the system through the three slits, just as shown in Fig.~\ref{FIG123}.

\subsection{Stronger-than-quantum correlations in time}

\begin{figure}\begin{center}
\includegraphics[width=14cm]{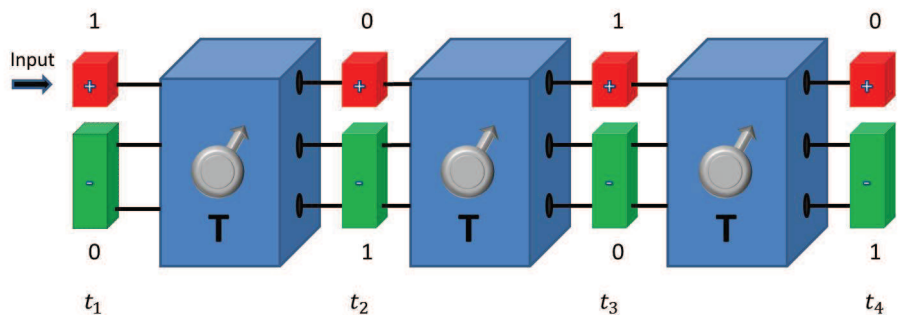}
\caption{Stronger than quantum temporal correlations.
The horizontal lines denote three possible paths a system can take, which can be thought to be represented by states $e_1$, $e_2$, and $e_3$ of the main. The evolution is driven by transformation $T$ of Eq.~(\ref{T}) and we define the dichotomic observable with outcomes $+1$ (red detector indicates system in state $e_1$) and $-1$ (green detector indicates system not in the state $e_1$). The numbers close to the detectors describe the probabilities of finding the system to have a particular outcome at various stages of evolution (if no measurement is performed at the earlier stage).
The system is measured successively at various pairs of time instances in order to establish two-point temporal correlations that enter into the Leggett-Garg inequality of Eq.~(\ref{LG}).
The evolution allows violation of the inequality more than what is permitted in the quantum theory (see main text).}
\label{FIG_LG}
\end{center}
\end{figure}

We now show that the theory of density cubes allows violation of the Leggett-Garg inequality beyond what is quantum mechanically possible.
The Leggett-Garg inequality involves temporal correlations between the measurement outcomes obtained at different instances of time.
Here we consider the measurement with a dichotomic
outcome $+1$ if the system is found in the state $e_1$, and outcome $-1$ if the system is not found in the state $e_1$, as shown in Fig.~\ref{FIG_LG}.

Consider a series of runs starting from identical initial conditions such that on the first set of runs the dichotomic observable $A$ is measured only at times $t_1$ and $t_2$,
only at $t_2$ and $t_3$ on the second set of runs, at $t_3$ and $t_4 $ on the third, and at $t_1$ and $t_4$ on the fourth ($0 < t_1 < t_2 < t_3 < t_4$).
Introducing temporal correlations $C_{ij}=\langle A(t_i)A(t_j)\rangle$ one can construct a combination of them in the form of the Clauser-Horne-Shimony-Holt expression
\begin{eqnarray}
K \equiv | C_{12} - C_{23} + C_{34} + C_{14} | \le 2,
\label{LG}
\end{eqnarray}
where the bound holds for classical-like theories (the Leggett-Garg inequality).
In quantum mechanics the bound depends on the degeneracy of the performed measurement~\cite{Budroni}. If the measurements are described by higher-rank projectors and the state follows the Luders update rule, quantum mechanics allows violation of this inequality up to $|K|_{QM}\leq2\sqrt2\approx 2.83$, the so-called Tsirelson bound for temporal correlations~\cite{Tsirelson,Fritz}.
However, within the framework of density cubes it can be readily verified that the scheme of Fig.~\ref{FIG_LG} predicts
$C_{12}=-1$, $C_{23}=0$, $C_{34}=-1$ and $C_{14}=-1$, and hence $K=3$.
Therefore, the experiments measuring temporal correlations and the strength of violation of the Leggett-Garg inequality can serve as tests of the cube theory.

\subsection{Higher-level systems}\label{HLSys}

In general the $N$-level system can be represented by a Hermitian cube $\rho_{ijk}$ where $i,j,k=1\dots N$.
The hermiticity condition implies that all the elements $\rho_{iij}$ and $\rho_{ijj}$ are real, and as it follows from the previous discussion are the ``quantum'' part of the state. Genuine non-quantum elements are those $\rho_{ijk}\in\mathbb{C}$ where all three indices are different and they define $2{N\choose 3}$ independent real parameters. In total a density cube of a $N$-level system has
\begin{eqnarray}
D(N) & = & N^2 -1 + 2{N\choose 3}
\end{eqnarray}
real parameters. The non-trivial dynamics can be generated by combining the operation $T$ defined in the previous section to different sets of three paths.

Note that the theory of density cubes violates the assumption of local tomography.
This assumption holds both in classical and quantum physics and asserts that the state of a composite system can be fully determined by combining data from measurements that determine the states of subsystems.
Therefore, the number of parameters $K(N_{AB})$ describing an unnormalised state of a composite system $AB$ satisfies $K(N_{AB})\leq K(N_{A})K(N_{B})$, where $K(N_{A})$ and $K(N_{B})$  are the number of real parameters required to describe an unnormalised state of systems $A$ and $B$~\cite{Hardy,Hardy&Wootters}. In the theory of density cubes this assumption is violated. As an example, consider the theory of density cubes for a four level system. The number of parameters reads $K(4)=D(4)+1=22>K(2)K(2)=(D(2)+1)(D(2)+1)=16$, so we are missing six parameters for the specification of an unnormalised state of a composite system.
Accordingly, these are exactly the non-quantum parameters that cannot be determined from local tomography and require joint measurements.

\section{State tomography}

Here we briefly give the procedure of reconstructing the density cube elements $\rho_{ijk}$ from the measurement data.
The quantum part of the cube can be reconstructed by means of standard quantum state tomography~\cite{Presti}.
The only non-trivial part is reconstruction of genuine non-quantum elements $\rho_{ijk}$ with all three indices being different.
For simplicity assume $N=3$ and therefore there is only one non-quantum complex element $\rho_{123}=z\in\mathbb{C}$.
%and let $\rho_{iii}=p_i\in[0,1]$. The probabilities $p_i$ can be obtained by measuring the state in the computational basis.
In order to extract $z$ element we apply the transformation $T$ of  equation (\ref{T}) and then apply the measurement in the standard basis.
Simple calculations show that the probabilities $\mu_i$ at the detectors are:
\begin{eqnarray}
\mu_1&=&\frac{1}{2}(p_2+p_3+2\sqrt3\mathrm{Re}z),\\
\mu_2&=&\frac{1}{2}(p_1+p_3-\sqrt3\mathrm{Re}z+3\mathrm{Im}z),\\
\mu_3&=&\frac{1}{2}(p_1+p_2-\sqrt3\mathrm{Re}z-3\mathrm{Im}z),
\end{eqnarray}
where $p_i = \rho_{iii}$ is the probability of measuring the $i$th result in the standard basis.
From here one can extract the value of $z$. Generalizations to higher dimensions are straightforward.

\section{Quantum interference and macroscopic realism}

Finally, we link second-order interference and violation of macroscopic realism, as introduced by Leggett and Garg~\cite{LG}. Similar ideas can be found in Ref.~\cite{Kofler}. We show that the simplest Leggett-Garg-type expression is of the form of $I_{12}$ term given in Eq. (\ref{I2}).
Under macro-realism, $I_{12} = 0$, and therefore macrorealism is violated by the phenomenon of quantum interference.
Similarly, the cube theory as well as all higher-order tensor theories are at variance with the premises of macroscopic realism.

Let us begin by recalling the assumptions of macro-realism~\cite{Legget}:
\begin{itemize}
  \item Macrorealism per se: ``A macroscopic object, which has available to it two or more macroscopically
  distinct states, is at any given time in a definite one of those states.''
  \item Noninvasive measurability: ``It is possible in principle to determine which of these states the system is in
  without any effect on the state itself, or on the subsequent system dynamics.''
\end{itemize}
Under these assumptions we now derive condition $I_{12} = 0$.
Consider an evolving macro-realistic system measured at times $t_1$ and $t_2$.
Assume that at time $t_1$ it is in one of two macroscopically distinct states.
One may think about a bullet shot into a double-slit experiment that at time $t_1$ propagates through one of the two slits.
Here the macroscopic state is the position of the bullet, i.e. at $t_1$ the position of the slit it goes through. The measuring apparatus at $t_1$ has the following four settings:
\begin{itemize}
  \item ``00'' Both slits open
  \item ``10'' Only second slit open
  \item ``01'' Only first slit open
  \item ``11'' Both slits closed
\end{itemize}
At time $t_2$, a different apparatus verifies if a system is in a possibly different macro-state we call ``$+$'' (this could be position of the object at the observation screen placed behind the screen containing the slits). Let us denote by $p_{ij}$ the probability of observing the ``$+$'' result at time $t_2$, if setting $ij$ was chosen at time $t_1$. Clearly, $p_{11} = 0$.
Due to the macro-realism assumptions the probability $p_{00}$ has to be given by the sum $p_{01} + p_{10}$,
as in every experimental run the system takes a well defined path through a single slit (macrorealism per se) and its future dynamics is independent of whether the unoccupied slit is opened or closed (non-invasiveness).
Therefore, condition $I_{12}=0$ can be seen as the simplest Leggett-Garg equality.

\section{Conclusions}

One might think that in order to observe genuine multi-slit interference it is necessary to modify the Born rule,
i.e. the probability of a particular result should not be given by the inner product between the relevant states of the theory but perhaps by its different power.
This is not the case as we presented here an explicit theory that does satisfy the Born rule but nevertheless allows for higher-order interference.
The state in the theory is represented by a mathematical object called ``density cube'' and is a natural generalization of the standard ``density matrix'' in quantum mechanics.
Quantum mechanics is contained in the theory of density cubes, but the latter in addition contains new ``coherence terms'' that give rise to the genuine third-order interference in Sorkin's classification.
The theory of density cubes is the first explicit example of a model exhibiting higher-order interference.
We have shown that density cubes allow violation of the temporal Tsirelson bound for the Leggett-Garg inequalities, thus illustrating the stronger-than-quantum correlations.
This result indicates an interesting relation between the Sorkin classification and the strength of correlations in the respective theories (see also~\cite{Niestegge}).

\ack

This research is supported by the National Research Foundation and Ministry of Education in Singapore.
TP acknowledges start-up grant of the Nanyang Technological University. This work was supported by the Austrian Science Fund (FWF) (Complex Quantum Systems (CoQuS), Special Research Program Foundations and Applications of Quantum Science (FoQuS), Individual project 24621), the European Commission  Project RAQUEL and by the Foundational Questions Institute (FQXi). We thank Gregor Weihs and Anton Zeilinger for discussion.

\end{document}